\documentclass{iaus}
\usepackage{graphicx}
\usepackage{natbib}

\title[Origin of CEMP stars] 
{What helium and lithium can tell us about CEMP stars?}

\author[Meynet et al.]   
{Georges Meynet$^1$,
Raphael Hirschi$^{2,3}$,
Sylvia Ekstrom$^1$,
Andr\'e Maeder$^1$,
Cyril Georgy$^1$,
Patrick Eggenberger$^1$,
Cristina Chiappini$^1$}

\affiliation{$^1$Geneva Observatory, Geneva University, \\ CH--1290 Sauverny, Switzerland \\ email: {\tt georges.meynet@unige.ch} \\[\affilskip]
$^2$Astrophysics group, Keele University, \\ Lennard-Jones Lab., Keele, ST5 5BG, UK 
\\ email: {\tt r.hirschi@epsam.keele.ac.uk }\\[\affilskip]
$^3$IPMU, University of Tokyo, \\ Kashiwa, Chiba 277-8582, Japan}

\pubyear{2010}
\volume{268}  
\pagerange{119--126}
\jname{Light elements in the Universe}
\editors{C. Charbonnel, M. Tosi, F. Primas, C. Chiappini, eds.}
\begin{document}

\maketitle

\begin{abstract}
We show that the peculiar surface abundance patterns of Carbon Enhanced Metal Poor (CEMP) stars has been inherited from
material having been processed by H- and He-burning phases in a previous generation of stars (hereafter called the ``Source Stars'').
In this previous generation, some mixing must have occurred between the He- and the H-burning regions in order
to explain the high observed abundances of nitrogen. In addition, 
it is necessary to postulate that a very small fraction of the carbon-oxygen core has been expelled (either by winds
or by the supernova explosion). Therefore only the most outer layers should have been released by the Source Stars.
Some of the CEMP stars may be He-rich if the matter from the Source Star is not too much diluted  with
the InterStellar Medium (ISM). Those stars formed from nearly pure ejecta would also be Li-poor.
\end{abstract}


\section{Introduction}

Observations have revealed in these last years a galactic halo much more complex than previously thought.
The existence of two halos, an inner one and an outer one \citep{Carollo2008} showed that, both from the point of view of kinematics and chemical composition, the galactic halo cannot be viewed as a single entity. At the smaller scale of the globular clusters, there are now ample evidences for the succession of at least two and maybe more stellar generations in clusters, each leaving a very peculiar imprint on the chemical composition of stars
(see the talks by Bragaglia and Decressin in the present volume and references therein). 
In that respect it is striking to note that stellar populations in clusters and in the field have each their own ``anomalous'' counterpart. In globular clusters the ``anomalous'' component  is made up of stars essentially made from material processed by only  H-burning reactions. In the field, the ``anomalous'' component is made up of stars presenting the nucleosynthetic signature of both H- and He-burning processes. 
In this paper, we focus our attention on the ``anomalous'' population observed in the field of the halo.  
In that population, we find the most iron poor star known today, HE 1327-2326, with a [Fe/H] as low as
-5.96 \citep{Frebel2008}. In this star, one counts 72 000 atoms of carbon for one atom of iron, while in the Sun, there is only 9 atoms of carbon per atom of iron!
Hence the name of  Carbon Enhanced Metal Poor Stars (CEMPs). The CEMP stars represent about 1 in 5 stars at metallicity below [Fe/H] $<$ -2.5  \citep{Lucatello2006}. 
In HE 1327-2326,  the numbers of nitrogen and oxygen atoms with respect to the number of iron atoms are respectively 20 000 times and 2500 times greater than in the Sun. Thus this star
could have also been named a N(itrogen)EMP or an Ox(ygen)EMP star! This star is either at the end of the Main-Sequence phase or just after the Main-Sequence (a subgiant), thus these abundances cannot have been 
produced in the star itself but must have been mainly inherited from the interstellar cloud from  which it formed about 13 billion years ago\footnote{Microscopic diffusion may
have altered the surface abundances however  \citep{Korn2009}, but this process cannot be responsible for the huge excesses in CNO elements.} or be acquired through a mass transfer episode in a close binary system. Whatever the scenario chosen,  the main question is what was the nature of the ``Source Stars'' which either enriched the natal cloud in such a peculiar way or
transferred its envelope to its less massive companion conferring it  its status of CEMP star? Was it a massive star? An intermediate mass star? Can we deduce some of its properties? 
This is the point we want to discuss in the present paper. 

\section{Constraints from nucleosynthesis}

\begin{figure}[t]
\includegraphics[width=2.5in,height=2.5in,angle=0]{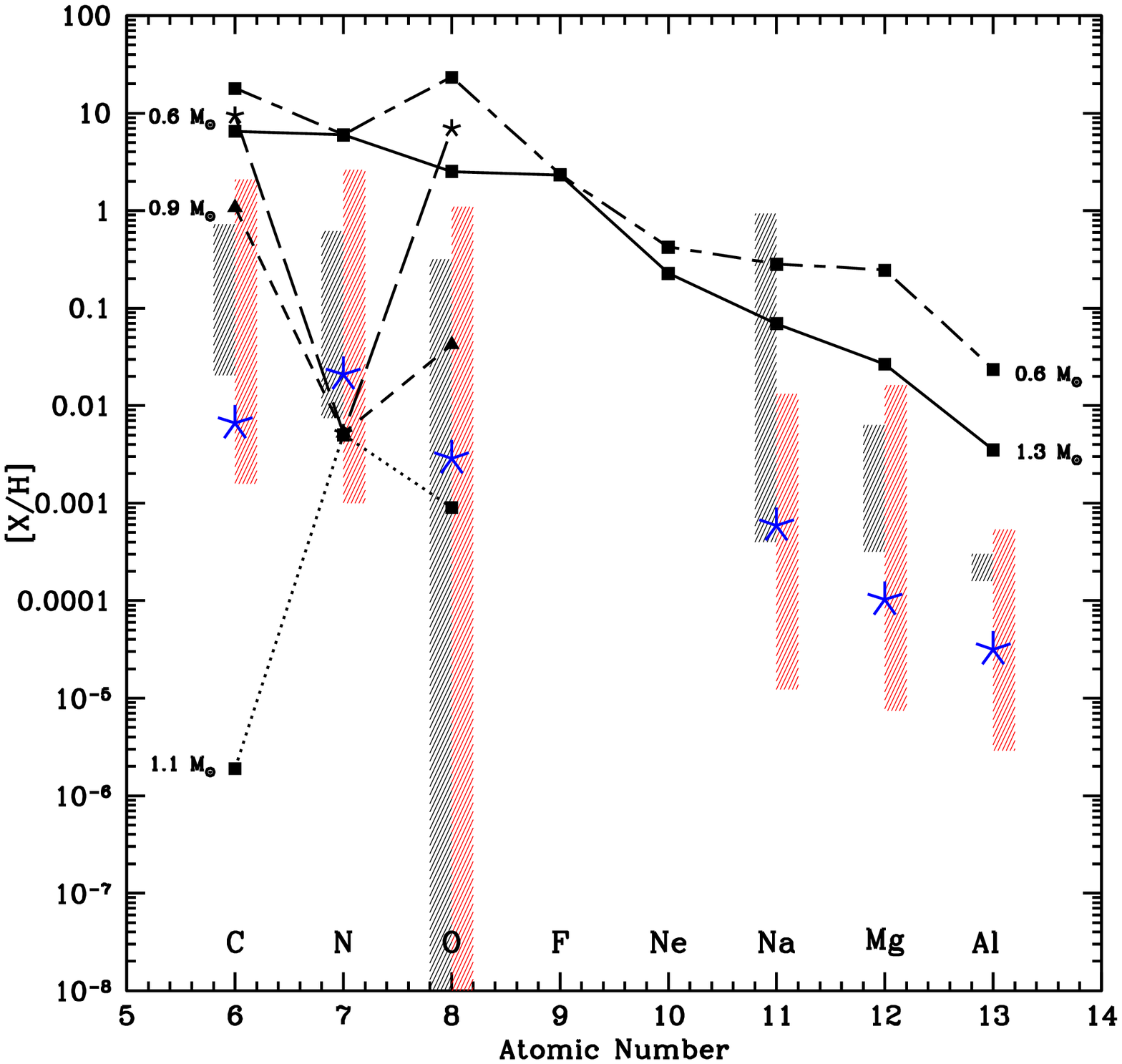}
\hfill
\hspace{1cm}
\includegraphics[width=2.5in,height=2.5in,angle=0]{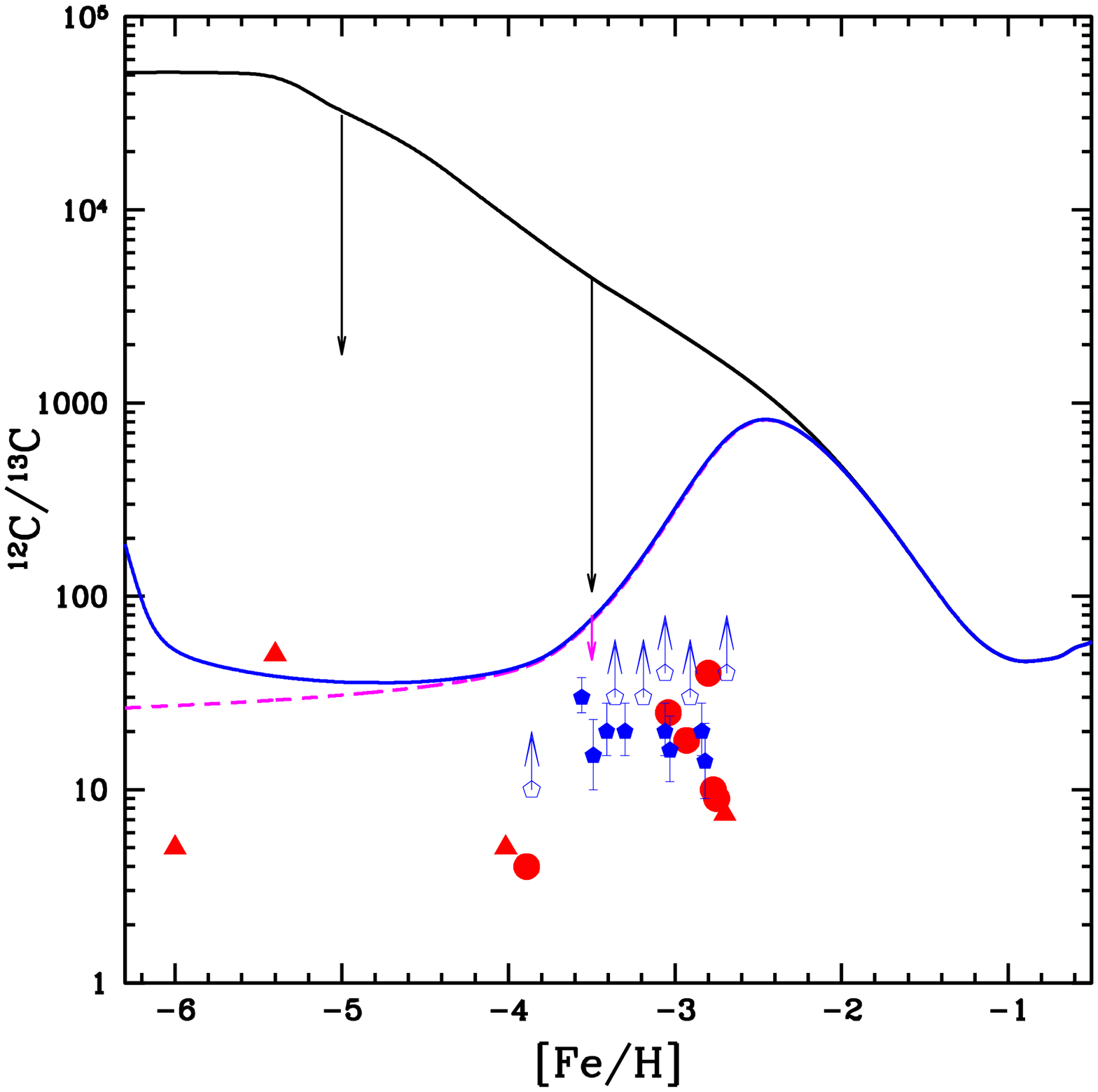}
\caption{{\it Left panel :}The continuous  and long-short dashed curves with filled squares show the
    composition obtained by mixing the outer  envelope above the langrangian mass coordinate 1.3 M$_\odot$ and 0.6 M$_\odot$  at the E-AGB phase of our Z=10$^{-5}$ rotating 7 M$_\odot$ stellar model.
    The dotted, short and long-dashed lines correspond to the composition of the envelope above the langrangian mass coordinate 1.1, 0.9  and 0.6 M$_\odot$  respectively at the E-AGB phase of our non-rotating 7 M$_\odot$ stellar model. 
    The CO core in the rotating and non rotating models have masses around 1.3 and 0.9 M$_\odot$ respectively.
     The star symbols show the observed values for the most iron poor star known today \citep{Frebel2008}.
     The vertical hatched zones show the range of observed values for CEMP-no stars \citep[i.e. those with
no evidence  of s-process element enhancements, see references in ][]{Masseron2009}. The dark left hatched zone corresponds to non-evolved stars, the left light grey zone to giants. {\it Right panel :}Predicted evolution of the
     $^{12}$C/$^{13}$C ratio according to chemical evolution models (CEM) for the halo computed with different stellar models for metallicities below Z=$10^{-5}$. The upper solid line corresponds to CEM models without
     fast rotators, the dashed and lower solid lines correspond to CEM models with fast rotators. The data are the unmixed stars of \citet{Spite2006}. The open symbols represent lower limits. The descending arrows indicate the final $^{12}$C/$^{13}$C ratio obtained after the first dredge-up in giants starting from the initial values given by the CEM models \citep[see details
     in][] {Chiappini2008}. The filled triangles (lower limits) and circles correspond to CEMP stars showing no s-process element enhancements and to the Frebel star (most iron poor star at [Fe/H] =-5.96).
}
\label{fig1}
\end{figure}

There are a few deductions that can be done without reference to any peculiar stellar models. These are the following:
\begin{itemize}
\item The CEMP stars present a great scatter in the abundances of the CNO elements (see the hatched zones in the left panel of Fig. \ref{fig1}). This is an indication that the chemical characteristics of  the material from which these stars inherited their peculiar composition  were different from case to case. This can be understood if the peculiar abundances reflect the chemical abundance of one or at a most a very small number of
nucleosynthetic events.
\item The very high overabundances with respect to iron of carbon, nitrogen and oxygen imply that the material responsible for these enhancements was mainly processed by the nuclear reactions involving nuclear reaction chains typical of  H- 
 (enhancement of N) and He-burnings (increases of C and O). In both nuclear phases, iron is not synthesized. Thus either iron has been produced in stars belonging to generations  having occurred before
the ``Source Stars'', or it has been produced, in part or in totality, by the ``Source Stars'' themselves in advanced nuclear phases. In that case,  the ``Source Stars'' should produce much smaller quantities than those
predicted by type II supernova explosions which are expected to produce [O/Fe] around 0.50 or less \citep[see for instance table 1 in ][]{Tominaga2007b}, while the Frebel star, for instance, show [O/Fe] of the order of 3.4!
\item The mixing of material having been processed by H-burning and of material processed by He-burning is however not a sufficient condition to reproduce the observed surface abundances of CEMP stars. To see what would be obtained by simple addition of material having been processed by these two nuclear phases {\it without allowing any mixing between them in the ``Source Star''}, we can look at the dotted, short and long dashed curves plotted in the left panel of Fig. \ref{fig1}. 
These curves show the chemical composition of the mixed outer envelope of a non-rotating 7 M$_\odot$ at Z=10$^{-5}$ at the early-AGB phase. 
Only the CNO elements are shown because these models were computed without the Ne-Na, Mg-Al chains.
When only the envelope
above 1.1 M$_\odot$ is ejected, we see matter processed only by H-burning mixed with some unaltered material in the outer envelope.
In case the initial metallicity of the model would have been chosen equal
to Z=10$^{-6}$ (maximum metallicity of stars which could have participated to the production of the material from which HE 1327-2326 could form), the whole curve would have been shifted downward by 1 dex. 
We see that material enriched in H-burning material cannot account for the observed  abundances in CEMP stars.  
When matter above 0.9 M$_\odot$ is ejected, some material processed by the He-burning reactions is added into the mixture. The abundance of nitrogen is not much changed, while those of carbon and oxygen increase a lot despite the fact that only
0.2 M$_\odot$ has been added with respect to previous case!
The trend is reinforced in case a still lower mass cut is considered. Considering a lower initial metal content, would in those cases, shift downward the nitrogen abundances without much changing the positions
of the carbon and oxygen abundances. We see thus that such models have no chance to reproduce the observed patterns. Similar conclusions can be reached looking at other initial mass models.
In order to achieve high abundances for all the three main CNO isotopes, nitrogen must also be produced from helium as carbon and oxygen. This is possible if
some carbon and oxygen produced in the helium core diffuse in the H-burning shell and are transformed there in nitrogen. {\it Thus the simultaneous large increases
of the three CNO elements are an indication that some mixing occurred between the He- and the H-burning regions}. 
\item Let us suppose that such a mixing occurred. The increase of nitrogen in the H-burning shell cannot be as high as the one of carbon and oxygen in the He-burning core. 
Diffusion is indeed not as efficient as convection and will never allow to homogenize carbon and oxygen between the He- and the H-burning regions. 
To see what happens when such a mixing occurs let us consider the left panel of Fig. \ref{fig1}. 
The upper continuous and long-short dashed curves show the abundances in the mixed outer envelope
of a rotating 7 M$_\odot$ stellar model with an initial value of $\Omega/\Omega_c$ equal to $\sim$ 0.8 where $\Omega$ is the surface angular velocity and $\Omega_c$ the critical angular velocity at the surface\footnote{The critical velocity is such that
when this velocity is reached, the centrifugal acceleration at the equator compensates for the gravity there.}. The model is in the early AGB phase. 
In such a model, during the whole
core He-burning phase, carbon and oxygen diffuse from 
the He-core into the H-burning shell.
We see that the envelope above the CO core (the CO core mass is about 1.3 M$_\odot$) is strongly enriched in CNO elements. Interestingly
it would also be strongly enriched in fluorine, neon (actually $^{22}$Ne), in $^{23}$Na, in magnesium (mainly $^{26}$Mg), and $^{27}$Al.  Mixing one
part of this envelope material from the ``Source Star'' with 100 parts of ISM would shift the curve downward by 2 dex providing a very good fit to the mean values of the observed patterns shown in that figure.
Since all these elements are formed either directly or indirectly from transformation of helium, the overabundances shown here are not so much dependant on the metallicity and would occur in a similar way in 
more metal poor stars. Also similar conclusions can be obtained from different initial mass models  \citep{Meynet2006, Hirschi2007}. However, depending on the initial mass of the model considered for the ``Source Star'',  the degree of dilution required to fit
the observed values can be different.
\end{itemize}

\section{The spinstar scenario}

In the ``spinstar scenario''
\citep{Meynet2006, Hirschi2007, Meynet2009iaurio}, the required mixing is induced by the instabilities triggered by  axial rotation of stars. Rotation can indeed induce mixing in radiative zones and thus allow
for instance carbon and oxygen to diffuse from the He core into the H-burning zone. This produces important amounts of nitrogen (primary nitrogen since it is produced from carbon and oxygen synthesized by the star itself)
and of primary $^{13}$C.  It is interesting to note that observations of the surface abundances of  ``normal'' metal poor halo field stars  (i.e. non CEMP stars) indicate that actually important amounts of primary nitrogen need to be produced in very metal poor massive stars. \citet{Chiappini2006, Chiappini2008} showed that rotating massive star models can reproduce the amount of primary nitrogen required by observation provided they began their
life with an initial angular momentum content of the same magnitude as the one in solar metallicity massive stars showing an averaged surface rotation rate during the MS of 200 km s$^{-1}$. In this model, the ``normal'' halo field stars are formed from a well mixed reservoir enriched by stars of different initial masses and of different initial metallicities. A consequence of this model is that low ratios of the
$^{12}$C/$^{13}$C should be observed at the surface of non-evolved metal poor halo field stars (see the right panel of Fig. 1). 

If now, instead of considering stars formed from material taken from  a well mixed reservoir, as above, we consider stars formed from material ejected by one single source, which kind of composition can we expect?
This is shown in the left panel of Fig. \ref{fig1} and was already discussed in the previous section. A good fit is obtained with the CEMP abundance patterns.
Thus we see that, in the frame of this model, the main difference between the ``normal'' and the ``anomalous'' population is not the nature of the ``Source Stars'' which in both cases
are rotating stars, but the fact that the normal populations are born from a well mixed reservoir, while the CEMP-stars are formed from a much more localized
and specifically enriched reservoir. 

An interesting feature of the present scenario is that rotational mixing can not only help in producing interesting abundance patterns, it can also be responsible for the loss through winds of the outermost layers.
Indeed, due to rotational mixing, the radiative envelope is enriched in CNO elements, its global opacity is increased and this may trigger line driven stellar winds, both in the case of massive stars in the supergiant phase and in the case
of intermediate mass stars along the AGB. Such an opacity increase only occurs when the stars have evolved beyond the Main-Sequence. Only at that time, can the surface be CNO enriched with values well above the initial metallicity.
This will only occur
at very low metallicity, let's say at metallicities below about 10$^{-5}$ . Why? For two reasons, first when the metallicity decreases rotational mixing is more efficient \citep{paperVII2001}, second the mass losses, both due to line radiation winds and to
mechanical winds due to
the reaching of the critical velocity are very weak at these low metallicities \citep[see][]{Meynet2009iaurio}, preventing the H-burning shell to disappear when the He-burning core is in activity. Thus only at low metallicities can we expect to
have winds enriched in both H- and He-burning products. At higher metallicities (10$^{-5} < $ Z $<10^{-3}$ ) , fast rotating models lose important amounts of matter through mechanical equatorial winds during the Main-Sequence \citep[][see also the contribution of Decressin in the present volume]{Decressin2007a}. This mechanical wind is enriched only in H-burning products. This may contribute to explain the origin of the different anomalous populations observed in the field and in the clusters of the halo: the clusters being more metal rich would be enriched by mechanical winds rich in H-burning products, while field halo stars may be enriched by line driven winds rich in both H- and He-burning products.

A prediction of the present scenario is that a very low $^{12}$C/$^{13}$C ratio can be obtained in CEMP stars. In the present models both isotopes are produced as primary elements, the
ratio is thus not much sensitive to the dilution factor (unless we consider so large dilution factors that it would also erase any strong overabundances in the CNO elements!). Very low
ratios are obtained in case the ``Source Star'' is a massive star having lost its outer envelope (below about 10). Higher ratios are obtained in the envelope of early AGB models \citep[of the order of 100, see Table 2 in][]{Meynet2009iaurio}. Thus this ratio
may be useful both for providing clues supporting the present ``spinstar'' scenario and also for disentangling between massive and intermediate mass stars (early AGB) as the possible ``Source Stars''. In the right panel of Fig. \ref{fig1}, we have
plotted the $^{12}$C/$^{13}$C ratios observed at the surface of the CEMP-no stars. We see that even with some CEMP  stars being dwarf stars (as for instance the filled circle at [Fe/H] equal to -4.015), the CEMP stars are in general below
the $^{12}$C/$^{13}$C ratio observed in giants (which should have lower  $^{12}$C/$^{13}$C ratios than dwarf stars since they went through the dredge-up episode). Moreover these CEMP stars are also below
the predictions of the chemical evolution model which describes the evolution of the composition of stars made up from the well mixed reservoir.
This two points support the view of a different origin of the CEMP star with respect to the normal halo stars based only from the $^{12}$C/$^{13}$C ratio, second it shows that
very low $^{12}$C/$^{13}$C ratios are observed in some dwarf CEMP star supporting, in view of the large $^{12}$C overabundances,
a primary origin for  $^{13}$C as predicted by the rotating models.
    
\section{Constraints from Li and He}   
   
Let us now come to the question asked in the title, namely how measures of the Li and He abundances in CEMP stars can be used to constrain the above scenario?
The material of the ``Source Stars'' are Li-poor (actually Li is completely destroyed) and He-rich \citep[see Table 2 in][]{Meynet2009iaurio}. Thus CEMP stars
made of pure ``Source Star'' material would be Li-poor and He-rich. Note that the same line of reasoning implies that stars presenting enhancements of Na and depletion 
of O in globular clusters should also be He-rich. 
Both in globular clusters and in the field, the actual level of He overabundance depends on the degree of dilution with interstellar medium \citep[see the contribution by Decressin in the present volume and references therein, see also][]{Maeder2006}.

Some Li observations exist for CEMP stars.  Many of the CEMP-no stars  show very low Li or even only upper limits compatible with no Li at all. Of course these
very low Li abundances can be due to depletion having occurred in the CEMP star itself and thus may not be the signature of a formation process from Li-depleted
material. A high helium abundance in the whole star (and not only at the surface) would be a much stronger constraint pointing towards a small dilution factor. 
The finding of such a star  would indeed be extremely interesting since in that case one would have an object providing a nearly direct insight
into the chemical composition of the ejecta of the first stars. Of course the difficulty here resides in how to measure helium abundance in a cool star. Although this is a real
challenge, the situation may not be completely desperate. A high helium abundance would imply a different initial mass associated to a given observed position in the HR diagram than the one obtained assuming
a normal (here a cosmological helium) abundance (see the contribution by Decressin in the present volume). 
The discovery of eclipsing binaries whose components would be CEMP stars would be a way to associate observationnally determined masses to given positions in the HR diagram and thus to test the He-rich hypothesis.
Asteroseismology would also be an interesting way to address that question. Finally, 
helium lines may be observable in cool stars,
whose strength might be related to the He abundance \citep{Moehler2000}.
While these possibilities appear quite exciting and will probably be explored in the future, it is however important to mention that
the absence of He-rich stars among the CEMP stars would not be an argument against the present spinstar scenario. High helium abundances
 imply small dilution factors (a factor of a few), while normal helium abundances imply larger dilution factors  (greater than an order of magnitude as in the case shown in the left panel of Fig. \ref{fig1}), but in both cases
 spinstars are required to explain the high CNO elements! It is also interesting to mention that the observed amounts of Li, Be and B produced by spallation reactions at very low metallicities give also some support
 to the present ``spinstar'' scenario  (see the review by Prantzos in this volume).

\bibliographystyle{aa}
\bibliography{MyBiblio}

\end{document}